\documentclass[]{article}       
\usepackage{emulateapj}
\usepackage{multicol}
\usepackage{psfig}
\lefthead{RADIO QUIET PULSARS}
\righthead{BARING \& HARDING}
\submitted{Astrophysical Journal Letters, in press}

\def\teq#1{$\, #1\,$}                           
\def\erg{\varepsilon}

\def\dover#1#2{\hbox{${{\displaystyle#1 \vphantom{(} }\over{
   \displaystyle #2 \vphantom{(} }}$}}

\def\today{\ifcase\month\or
  January\or February\or March\or April\or May\or June\or
  July\or August\or September\or October\or November\or
  December\fi
  \space\number\day, \number\year}

\begin{document}

\newcommand{\vol}[2]{$\,$\rm #1\rm , #2.}                 
\newcommand{\figureout}[2]{ \figcaption[#1]{#2} }       

\title{RADIO QUIET PULSARS WITH ULTRA-STRONG MAGNETIC FIELDS}
\author{Matthew G. Baring\altaffilmark{1} \& Alice K. Harding}
\affil{Laboratory for High Energy Astrophysics \\
       Code 661, NASA/Goddard Space Flight Center \\
       Greenbelt, MD 20771, U.S.A.\\
      \it baring@lheavx.gsfc.nasa.gov, harding@twinkie.gsfc.nasa.gov\rm}
   \altaffiltext{1}{Universities Space Research Association}
\date{\today}

\begin{abstract} 
The notable absence of radio pulsars having measured magnetic dipole
surface field strengths above \teq{B_0\sim 3\times 10^{13}} Gauss
naturally raises the question of whether this forms an upper limit to
pulsar magnetization.  Recently there has been increasing evidence that
neutron stars possessing higher dipole spin-down fields do in fact
exist, including a growing list of anomalous X-ray pulsars (AXPs) with
long periods and spinning down with high period derivatives, implying
surface fields of \teq{10^{14}}--\teq{10^{15}} Gauss.  Furthermore, the
recently reported X-ray period and period derivative for the Soft
Gamma-ray Repeater (SGR) source SGR1806-20 suggest a surface field
around \teq{10^{15}} Gauss.  None of these high-field pulsars have yet
been detected as radio pulsars.  We propose that high-field pulsars
should be radio-quiet because electron-positron pair production in
their magnetospheres, thought to be essential for radio emission, is
efficiently suppressed in ultra-strong fields (\teq{B_0\gtrsim 4\times
10^{13}}Gauss) by the action of photon splitting, a quantum
electrodynamical process in which a photon splits into two.  Our
computed radio quiescence boundary in the radio pulsar \teq{P-\dot P}
diagram, where photon splitting overtakes pair creation, is located
just above the boundary of the known radio pulsar population, neatly
dividing them from the AXPs.  We thus identify a physical mechanism
that defines a new class of high-field radio-quiet neutron stars that
should be detectable by their pulsed emission at X-ray and perhaps
$\gamma$-ray energies.
\end{abstract}


\section{INTRODUCTION}
   \label{introduction}

The magnetic fields of pulsars have traditionally been inferred through
measurement of both a period \teq{P} and a period derivative \teq{\dot
P}.  Presuming that the increase in period results from spin-down of
the neutron star due to electromagnetic dipole radiation energy loss,
the derived surface field is (Shapiro \& Teukolsky 1983; Usov \&
Melrose 1995) \teq{B_0 = 6.4 \times 10^{19}\, (P \dot{P})^{1/2}}Gauss;
this estimate assumes a braking index of 3, and a uniformly magnetized
stellar interior as opposed to the Manchester and Taylor (1977)
convention of a pure dipole field.  A plot of \teq{P} vs.  \teq{\dot P}
for radio pulsars with measured \teq{\dot P} is shown in
Figure~\ref{fig:quiescence}.  The derived surface fields of these
pulsars spans a range of five decades from \teq{3 \times 10^8} Gauss to
\teq{3 \times 10^{13}} Gauss, as indicated by the diagonal lines of
constant \teq{B_0}.  No radio pulsars have inferred fields above this
range, even though known selection effects for a given pulsar age do
not {\it a priori} prevent their detection (Cordes, private
communication).  Also shown in Figure~\ref{fig:quiescence} is the
so-called ``death line,'' with \teq{\dot P P^{-3}=\hbox{const.}} (i.e.
for fixed open field line voltage), to the right of which there are no
detected pulsars.  The absence of radio pulsars at periods larger than
5s has been explained as a turn-off in the radio emission when the
$\gamma$-rays emitted by particles accelerated near the polar cap
cannot produce electron-positron pairs (Sturrock 1971; Ruderman \&
Sutherland 1975; Arons 1983).

In addition to the radio pulsars, we have plotted in
Figure~\ref{fig:quiescence} the AXPs having measured \teq{P} and
\teq{\dot P}.  The AXPs are a group of six or seven pulsating X-ray
sources with periods around 6-12 seconds, which are anomalous in
comparison with average characteristics of known accreting X-ray
pulsars.  They are bright, steady X-ray sources having luminosities
\teq{L_X \sim 10^{35}\,\rm erg\; s^{-1}}, they show no sign of any
companion, are steadily spinning down, and have ages \teq{\tau \lesssim
10^5} years.  Those which have measured \teq{\dot P} (e.g. Mereghetti
\& Stella 1995; Gotthelf \& Vasisht 1998) are shown in Table~1 and have
derived magnetic fields between \teq{10^{14}} and \teq{10^{15}} Gauss,
assuming that they are spinning down due to dipole radiation torques.
The soft gamma-ray repeaters (SGRs) are another type of high-energy
source that has recently joined this group of highly magnetized neutron
stars.  SGRs are transient $\gamma$-ray burst sources that undergo
repeated outbursts.  There are four known SGR sources (one newly
discovered: Kouveliotou, et al. 1998b), with two (SGRs 1806-20 and
0525-66) associated with young (\teq{\tau < 10^5} yr) supernova
remnants.  Recently, 7.47s and 5.16s pulsations have been discovered
(Kouveliotou, et al. 1998a,c; Hurley et al. 1998) in the quiescent
X-ray emission of SGRs 1806-20 and 1900+14, respectively, with SGR
1900+14 exhibiting a 5.15s period in a $\gamma$-ray burst (Cline
et al. 1998).  In the \teq{P-\dot P} diagram, both AXPs and SGRs live
in a separate region above the detected radio pulsars.  Such a class of
highly magnetized neutron stars, or ``magnetars", was postulated as a
model for SGRs (Thompson \& Duncan 1993) and more recently also for the
AXPs (Thompson \& Duncan 1996).  According to the model, magnetars are
neutron stars born with millisecond periods that generate magnetic
fields above \teq{10^{14}} Gauss by dynamo action due to convective
turbulence.


\newpage

\centerline{}
\vskip 0.2truecm
\centerline{\psfig{figure=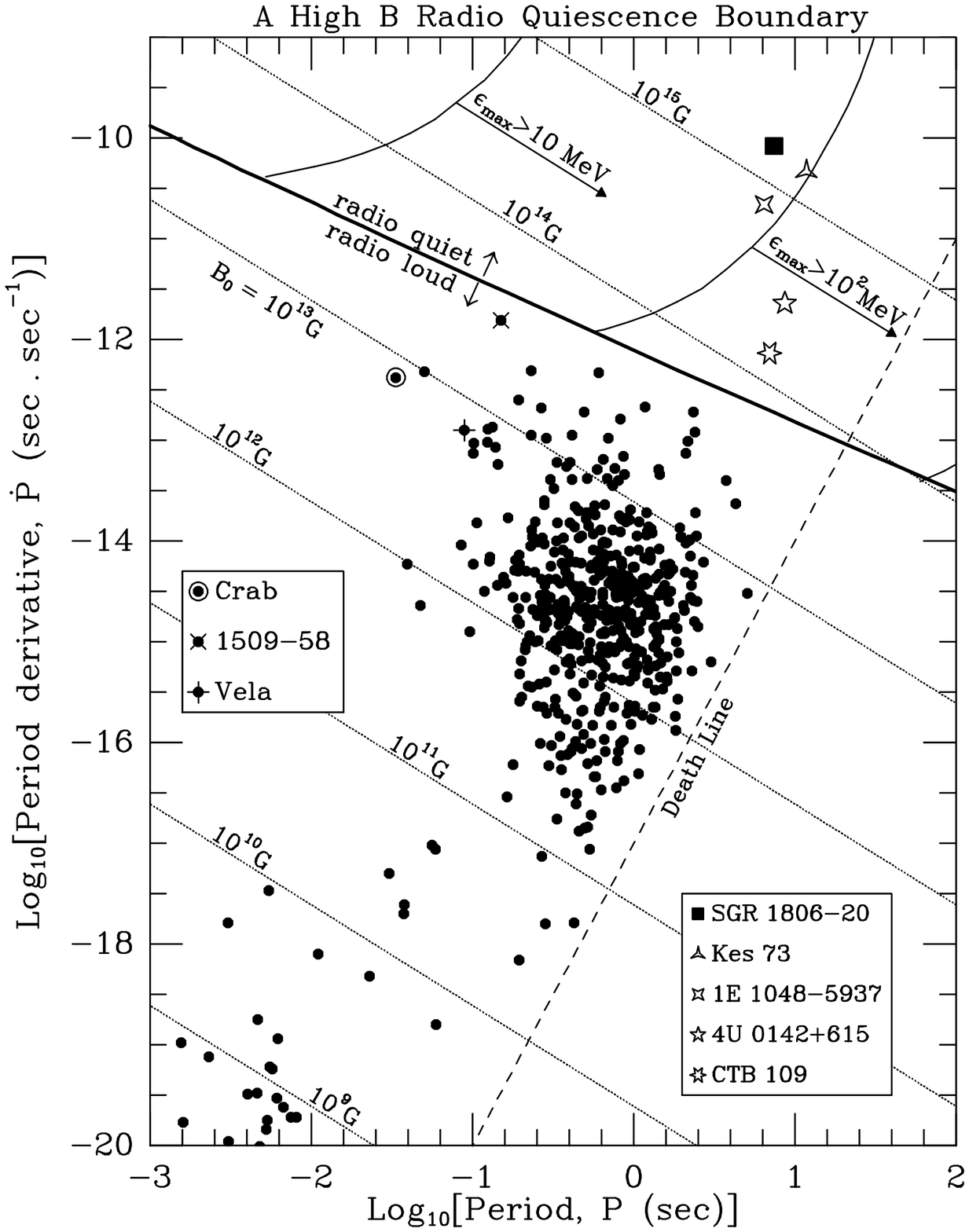,width=8.8cm}}
\vskip 0.0truecm
\figcaption{
The conventional depiction of pulsar phase space, the
\teq{P}-\teq{\dot{P}} diagram, with filled circles denoting the pulsars
listed in the Princeton Pulsar Catalogue (Taylor, Manchester \& Lyne
1995; see also {\tt http://pulsar.princeton.edu/}), together with their
choice of a fiducial position for the long period death line.  Three
gamma-ray pulsars are highlighted, labelled as in the inset.  The heavy
curve denotes the approximate boundary of radio quiescence for photons
initially propagating along the field, below which pulsars can be radio
loud, and above which photon splitting suppresses pair creation so that
pulsars should be radio quiet.  Photons were assumed to emanate from
near the stellar surface.  Also shown on the upper right are the
positions of four anomalous X-ray pulsars and SGR 1806-20, whose
measured \teq{\dot P} and inferred fields \teq{B_0} are listed in
Table~1.  The curves marked \teq{\erg_{\rm max}>10}MeV and
\teq{>100}MeV are contours (drawn only in the radio quiet region) for
the photon splitting escape energy (see text); to the right of these
curves the magnetosphere is transparent to 10 MeV and 100 MeV photons,
respectively.
   \label{fig:quiescence} }       
\centerline{}

This paper argues that there is a simple physical reason (within the
confines of polar cap surface emission models) why radio emission
should be absent in pulsars with ultra-strong fields.  The behaviour of
quantum mechanical radiation processes in magnetar-type fields is
distinctly different from that of conventional radio pulsars, where
\teq{B_0\lesssim 10^{13}}Gauss.  While the purely quantum process of
single-photon pair production \teq{\gamma\to e^+e^-} is commonly
invoked in radio pulsar models (following Sturrock 1971) of sources
such as the Crab and Vela, at the high field strengths appearing in
magnetars, the more exotic process of photon splitting,
\teq{\gamma\to\gamma\gamma}, can become quite probable.  The splitting
of one photon into two lower energy ones (the rate for which was first
calculated correctly by Adler et al. 1970; Bialynicka-Birula \&
Bialynicki-Birula 1970; and Adler 1971), which is third-order (in the
fine structure constant) in quantum electrodynamics (QED), can act as a
competitor to pair creation \teq{\gamma\to e^+e^-} as an attenuation
mechanism for gamma-rays.  Photon splitting is also a purely quantum
effect, and has appreciable reaction rates (Adler 1971; Baring \&
Harding 1997) only when the magnetic field is at least a significant
fraction of the quantum critical field \teq{B_{\rm cr}=4.413\times
10^{13}}Gauss; it forms the centerpiece of the discussion here due to
its potential for suppressing the creation of pairs.  Presuming that a
plentiful supply of pairs is a prerequisite for, and maybe also
guarantees, coherent radio emission at observable levels (a premise of
standard polar cap models for radio pulsars: e.g. Sturrock 1971;
Ruderman \& Sutherland 1975; Arons 1983), an immediate consequence of
the suppression of pair creation in pulsars is that detectable radio
fluxes should be strongly inhibited.  We propose here that when
splitting \teq{\gamma\to\gamma\gamma} dominates the attenuation of
gamma-rays, pulsars should be radio quiet.

\centerline{}
\vskip 0.0truecm
\centerline{\psfig{figure=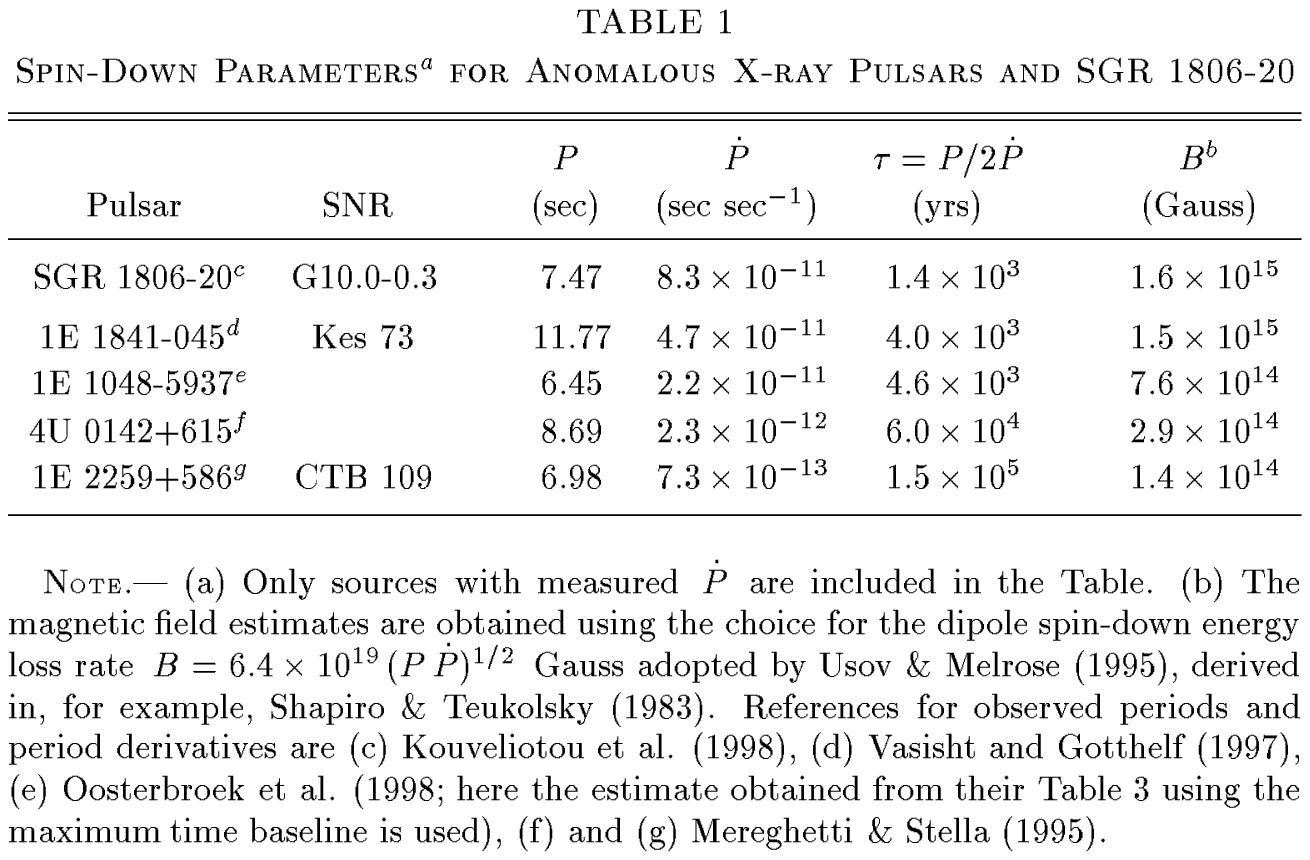,width=8.8cm}}
\vskip 0.0truecm
\centerline{}

\section{CRITERION FOR RADIO QUIESCENCE}
 \label{quiescence}

The determination of the pulsar parameter space (i.e.  \teq{B_0},
\teq{P}, \teq{\dot P}) when such a suppression of pairs by photon
splitting becomes efficient is therefore of paramount interest.  To do
this, one must compare the attenuation probabilities of photons by the
two quantum processes, \teq{\gamma\to e^+e^-} and
\teq{\gamma\to\gamma\gamma}, during their passage through the pulsar
magnetosphere from emission points near the stellar surface.  Polar cap
pulsar models invoke an electric field \teq{E_{\parallel}} parallel to
{\bf B} that rapidly accelerates electrons and/or positrons to Lorentz
factors of around \teq{10^5}--\teq{10^7} (e.g. Ruderman \& Sutherland
1975; Arons 1983; Usov \& Melrose 1995; Harding \& Muslimov 1998).  If
the polar field \teq{B_0} is not too high (Harding \& Muslimov 1998),
these energetic particles radiate curvature photons (Daugherty \&
Harding 1982), while resonant Compton scattering of quasi-thermal
X-rays from the stellar surface is the dominant radiation production
mechanism (Sturner 1995; Harding \& Muslimov 1998) when \teq{B\gtrsim
10^{13}} Gauss.  These emission processes beam the primary radiation
into a very narrow cone virtually aligned with the local magnetic
field.  Non-zero attenuation coefficients \teq{T(\omega, \, \theta_{\rm
kB} )} (defined as the rate divided by \teq{c}) for \teq{\gamma\to
e^+e^-} and \teq{\gamma\to\gamma\gamma} are only possible for oblique
angles \teq{\theta_{\rm kB}} of photon propagation with respect to the
field: this can be deduced from the Lorentz transformation property
(Baring \& Harding 1998) \teq{T(\omega, \, \theta_{\rm
kB})=\sin\theta_{\rm kB} T(\omega \sin\theta_{\rm kB},\,\pi/2 )} for
photons of energy \teq{\omega m_ec^2}.  Hence, to obtain significant
optical depths to attenuation, propagation of radiation through the
magnetosphere is essential to effect crossing of field lines.

Both \teq{\sin\theta_{\rm kB}} and \teq{\omega \sin\theta_{\rm kB}}
generally increase monotonically during propagation away from the polar
cap (Harding, Baring \& Gonthier 1997).  Pair creation cannot take
place before its threshold of \teq{2m_ec^2/\sin\theta_{\rm kB}} is
crossed during such propagation.  Even though photon splitting has a
lower rate (being a higher-order QED process), it may dominate pair
creation because it has no threshold.  This effect is illustrated in
Fig.~1 of Baring \& Harding (1997), which also exhibits clearly that
\teq{T(\omega\sin\theta_{\rm kB},\,\pi/2 )} is a strongly increasing
function of both \teq{\omega\sin\theta_{\rm kB}} and \teq{B} for both
pair creation and photon splitting.

Computing attenuation probabilities of the two processes for photon
propagation near a neutron star requires the incorporation of general
relativistic effects (Harding, Baring \& Gonthier 1997; Baring \&
Harding 1998).  Curved spacetime modifies photon trajectories from
rectilinear propagation and therefore changes the values of
\teq{\theta_{\rm kB}}, redshifts photon energies, and increases the
magnetic field in the local inertial frame.  All three of these effects
can lead to significant reductions in the attenuation lengths of the
two quantum processes.  Details of such attenuation calculations are
deferred to Baring \& Harding (1998); here we focus on their
consequences.  The first of these is that for both processes, due to
the \teq{r^{-3}} falloff of the dipole field, for each photon emission
point there always exists (Harding, Baring \& Gonthier 1997) an energy
called {\it the escape energy}, below which the magnetosphere is
transparent to the radiation, and above which it is opaque and photon
escape is prohibited.  Such domains of opacity at high energies are a
principal reason why the polar cap model predicts that pulsars like the
Crab and Vela cannot be visible above a maximum energy \teq{\erg_{\rm
max}} of a few GeV (Daugherty \& Harding 1982), and perhaps (Harding,
Baring \& Gonthier 1997) also why no emission is observed for
PSR1509-58 above a few MeV.

An approximate measure of when the creation of pairs is strongly
suppressed by photon splitting is when their {\it escape energies
\teq{\erg_{\rm esc}} are equal}.  A more accurate measure will require
detailed simulation of splitting/pair cascades.  Generally in fields
surmised for Crab-like and Vela-like pulsars, \teq{\erg_{\rm esc}} for
splitting exceeds that for \teq{\gamma\to e^+e^-} so that photon
attenuation only takes place via conversions into pairs.  In near
critical and supercritical fields (\teq{B\gtrsim B_{\rm cr}}), the
opposite occurs (Harding, Baring \& Gonthier 1997; Baring \& Harding
1998) and splitting assumes the role of the dominant attenuation
mechanism, strongly (but not totally) suppressing the creation of
pairs.  This effect appears to be inevitable, regardless of whether
free pair creation or positronium formation is considered (Baring \&
Harding 1998), given the polarization properties of the photon
production mechanisms of curvature and synchrotron radiation, and
resonant Compton scattering, and those of the two attenuation
processes.

The condition of equality of escape energies for splitting and pair
creation therefore implicitly defines an approximate {\it boundary of
radio quiescence}, which for a given colatitude \teq{\Theta} for the
emission of a photon near the surface, corresponds (Baring \& Harding
1998) to a particular value of the spin-down dipole field \teq{B_0}.
Such a \teq{(B_0,\,\Theta )} relationship can be quickly transformed
into one between the principal radio pulsar observables, namely
\teq{P}, the pulsar period (in seconds), and \teq{\dot{P}}, the period
derivative.  This is effected using the standard relationship
\teq{\sin\Theta =[2\pi R/(Pc)]^{1/2}} between the period and the polar
cap size or colatitude \teq{\Theta} for a neutron star of radius
\teq{R}, modified (Baring \& Harding 1998) by geometric distortions
introduced by curved spacetime, and the dipole radiation formula
(Shapiro \& Teukolsky 1983; Usov \& Melrose 1995) \teq{B_0 = 6.4\times
10^{19}\, (P \dot{P})^{1/2}}Gauss.  The boundary, thus transformed, is
depicted in Figure~\ref{fig:quiescence} in the conventional
\teq{P}-\teq{\dot{P}} diagram for the radio pulsar population, and
corresponds roughly to 
\begin{equation}
   \dot{P}\;\approx\; 7.9\times 10^{-13}\; 
   \biggl(\dover{P}{1\,\hbox{sec}}\biggr)^{-11/15}\;\; .
 \label{eq:deathline}
\end{equation}
The particular value \teq{-11/15} of the index follows from the
dependence of the rate of photon splitting on energy (\teq{\propto
\erg^5}), photon angles to the field (\teq{\propto\theta_{\rm kB}^6}),
and the magnetic field in this regime (approximately proportional to
\teq{B_0^{15/4}} for \teq{B_{\rm cr}\lesssim B_0\lesssim 4B_{\rm cr}}),
coupled with the fact that propagation guarantees that \teq{\theta_{\rm
kB}\propto\Theta}, and that the pair creation escape energy scales as
\teq{\Theta^{-1}} (see Baring \& Harding 1998).  This putative boundary
for radio quiescence, which is computed specifically for photon origin
near the stellar surface, {\it is comfortably located above the entire
collection of radio pulsars}, a situation that may be tested by the
imminent dramatic increase in the population due to the new multi-beam
Parkes survey (Manchester 1998).  A study of self-consistent particle
acceleration above a pulsar polar cap reveals that emission near the
neutron star surface is expected for pulsars having \teq{B\gtrsim
10^{13}} Gauss (Harding \& Muslimov 1998), principally because the pair
formation fronts are established in regimes where resonant Compton
scattering dominates the cooling of primary electrons and created
pairs.

Above the boundary, pair creation is strongly inhibited and by
extension, pulsars should be radio quiet.  The absence of radio pulsars
in this high-\teq{B_0} region of phase space has become of much greater
importance with the mounting observational evidence for neutron stars
with such large fields.  This evidence includes the detection of
spin-down in the growing number of anomalous X-ray pulsars, which when
combined with their association with young supernova remnants, has
shifted the focus from accretion torques (e.g. Mereghetti \& Stella
1995; van Paradijs, Taam \& van den Heuvel 1995) to electromagnetic
dipole torques (e.g. Vasisht \& Gotthelf 1997) as the origin of the
spin-down, implying immense supercritical (\teq{B_0> 4.41\times
10^{13}}Gauss) fields in these sources.  These are radio quiet pulsars,
as is the quiescent counterpart of SGR 1806-20, whose 7.47 second
periodicity has only been very recently announced (Kouveliotou, et al.
1998a).  The spin-down parameters for these sources, listed in Table 1,
are used to derive their positions on the \teq{P}--\teq{\dot P} diagram
in Figure~\ref{fig:quiescence}.  Our approximate boundary for radio
quiescence neatly separates the radio loud and radio quiet pulsar
populations.  We contend that this pulsar dichotomy is a simple
consequence of the quantum physics in strong fields discussed herein.

\newpage

\section{DISCUSSION}
 \label{discussion}

The location of the radio quiescence boundary above the detected radio
pulsar population gives consistency with the assumption of high-energy
photon emission at the rim of a standard polar cap: larger cap sizes
would permit photons to acquire large angles to the field more quickly,
pushing the quiescence boundary to lower \teq{B_0}, in conflict with
the data.  In the extreme case where the emission does not arise at the
polar caps but occurs in equatorial regions of the magnetosphere, as
advocated (Baring \& Harding 1995) in SGR models, photon splitting can
dominate even more efficiently.  There are a number of other reasons
why the emission properties of ultra-magnetized pulsars far above the
radio quiescence line may well be quite different.  The fact that the
observed luminosities of AXPs are much larger than their spin-down
luminosities requires an alternative energy source to power their
emission.  Decay of the magnetic field, predicted to be much faster in
magnetars than in radio pulsars (Thompson \& Duncan 1996), may heat the
neutron star surface and generate Alfv\'en waves that lead to particle
acceleration and non-thermal emission in the magnetosphere.  Since the
energy density of these fields produces a high level of stress on the
neutron star crust, continual glitches and crustal fracturing may
occur.  The emission from these sources will be governed by QED physics
of ultra-strong fields, making them unique laboratories for the study
of exotic processes such as photon splitting and even trident
production, neutrino and muon pair production, or magnetic \v{C}erenkov
radiation, in addition to more commonplace mechanisms such as
cyclotron/synchrotron radiation and resonant Compton scattering.

The observational expectations for high-field pulsar emission in the
radio, X-ray and gamma-ray bands can be quickly summarized in view of
the physical effect of suppression of pair creation.  Clearly, radio
luminosity should decline as the boundary of radio quiescence is
approached from below, an effect that can be enhanced by the creation
of pairs in low Landau levels (Harding \& Daugherty 1983) or
positronium formation (Usov \& Melrose 1995) above \teq{B_0\gtrsim
6\times 10^{12}} Gauss.  Note that photo-ionization and electric
field dissociation of positronium can mute the effectiveness of bound
state pair creation in inhibiting radio emission (such issues are
discussed in Baring \& Harding 1998).  Since synchrotron radiation, the
dominant emission of produced pairs, has a steeper spectrum than
primary curvature or resonant Compton-scattered radiation, the absence
of a synchrotron component should lead to a progressive flattening of
pulsar gamma-ray spectra as \teq{B_0} is increased to values above
\teq{\sim 4\times 10^{13}}Gauss.  In the magnetar regime of
ultra-strong fields, the maximum energy of emission \teq{\erg_{\rm
max}} is dictated by the escape energy \teq{\erg_{\rm esc}} for
splitting; sample values are illustrated in
Figure~\ref{fig:quiescence}, indicating that SGR1806-20 in its
quiescent state and the Kes 73 and 1E 1048-5937 AXPs would barely be
visible at around 100 MeV.  A decline in gamma-ray luminosity below the
\teq{B_0/P^2} proportionality found for conventional gamma-ray pulsars
(e.g. Thompson, et al. 1997) would also be expected.  In the X-ray
range, non-thermal spectra would probably also flatten as \teq{B_0}
increased above \teq{10^{13}}Gauss, as the cyclotron energy moved into
the MeV range.  However, at much higher fields, the AXPs exhibit steep
X-ray continuum emission (and probably distinct from any hard X-ray or
gamma-ray component), at luminosities far exceeding that available due
to dipole spin-down.  Hence, given that another mechanism for energy
supply such as field decay (Thompson \& Duncan 1996) must be operating
in these sources, how their spectrum and luminosity in the X-ray band
would depend on \teq{B_0} and \teq{P} is not well understood.

The prediction of a radio quiescence boundary, coupled with the
expectations mentioned just above for emission properties in the soft
gamma-ray band, provide strong observational tests for the quantum
effects in strong magnetic fields relating to pair production and its
suppression by photon splitting.  Future studies by radio telescopes
and measurements by X-ray experiments such as AXAF and XMM, and
gamma-ray missions such as Integral and GLAST should delve deeper into
the exciting phenomenon of pulsars with ultra-strong magnetic fields.

\acknowledgments
We thank Alex Muslimov, Jim Cordes and Eric Gotthelf for discussions.
MGB acknowledges the support of a Compton Fellowship during the period
when part of this research was completed.  This work was supported by
the NASA Astrophysics Theory Program.

\end{document}